\documentclass[aps,pre,reprint,longbibliography,floatfix]{revtex4-2}
\usepackage[pdftex]{graphicx}
\usepackage{float}
\usepackage[caption=false]{subfig}

\begin{document}
\title{Heartbeat instability as auto-oscillation between dim and bright void regimes}
\author{A. Pikalev}
\author{M. Pustylnik}
\author{C. R\"ath}
\author{H. M. Thomas}
\affiliation{Institut f\"ur Materialphysik im Weltraum, Deutsches Zentrum f\"ur Luft- und Raumfahrt e.\,V. (DLR), 82234 We{\ss}ling, Germany}
\email{Aleksandr.Pikalev@dlr.de}

\begin{abstract}
We investigated the self-excited as well as optogalvanically stimulated heartbeat instability in RF discharge complex plasma.
Three video cameras measured the motion of the microparticles, the plasma emission, and the laser-induced fluorescence simultaneously.
Comprehensive studies of the optogalvanic control of the heartbeat instability revealed
that the microparticle suspension can be stabilized by a continuous laser, whereas a modulated laser beam induces the void contraction either transiently or resonantly.
The resonance occurred when the laser modulation frequency coincided with the frequency of small breathing oscillations of the microparticle suspension,
which are known to be a prerequisite to the heartbeat instability.
Based on the experimental results we suggest that the void contraction during the instability is caused by an abrupt void transition
from the dim to the bright regime [Pikalev et al., Plasma Sources Sci. Technol. \textbf{30} 035014 (2021)]. 
In the bright regime, a time-averaged electric field at the void boundary heats the electrons causing bright plasma emission inside the void.
The dim void has much lower electric field at the boundary and exhibits therefore no emission feature associated with it.

A.~Pikalev et~al. {\em Phys.~Rev.~E} \textbf{104}, 045212, (2021). DOI:
\href{https://doi.org/10.1103/PhysRevE.104.045212}{10.1103/PhysRevE.104.045212}.

\copyright\,2021 American Physical Society.
\end{abstract}

\maketitle

\section{INTRODUCTION}

Complex or dusty plasma is a medium containing ionized gas and micron-sized solid particles \cite{Fortov-PhysRep2005, Morfill-RevModPhys2009, Ivlev2012}.
The strongly coupled subsystem of such microparticles can be used 
for particle resolved studies of generic classical phenomena in condensed matter.
For such studies, homogeneous microparticle suspensions are desirable. 
However, even microgravity does not guarantee homogeneous distribution and stability of the microparticle component. 
A void, i.~e. a microparticle-free area, is one of most common disturbances of the complex plasma homogeneity under microgravity conditions.
The void formation and growth also determine the nanoparticle generation cycle in plasma reactors \cite{Pilch-JApplPhys2017}.

Due to its small charge-to-mass ratio and its mesoscopic nature, the dust component introduces a variety of instabilities 
including dust density waves \cite{Merlino-JPlasPhys2014},
transverse instability \cite{Zobnin-JPhysConfSeries2016},
filamentary mode and the void rotation \cite{Praburam-AIPPhysPlas1996, Samsonov-PhysRevE1999, Mikikian-AIPPhysPlas2006},
plasmoids and carousel instability \cite{Mikikian-PhysRevLett2012, Mikikian-IEEETPS2011}.
Being multi-timescale phenomena, the instabilities in complex plasma pose a hard problem for numerical studies.
The physical mechanism of most of these instabilities is still unknown.

In the first microgravity experiments with complex plasmas, a spontaneous periodic contraction of the void boundary was reported \cite{Goree-MFPTP1998}.
Because of its characteristic appearance as well as due to very low repetition frequency 
(from single contractions to hundreds Hz), this phenomenon was termed ``heartbeat'' instability.
This particular instability is the subject of the present work.

The heartbeat instability is a complex auto-oscillation phenomenon,
which was treated in \cite{Mikikian-PhysRevLett2008} as mixed-mode oscillations.
Mixed-mode oscillations occur when a dynamical system switches between fast and slow motion and small and large amplitude \cite{Brons-Chaos2008}.
The mixed-mode oscillations are encountered in a wide range of phenomena including glow discharges \cite{Braun-PhysRevLett1992, Hayashi-PhysRevLett2000},
chemical processes \cite{Albahadily-JChemPhys1989, Petrov-JChemPhys1992}, 
and models of neuronal activity \cite{Rubin-BioCyber2007} and the real mammal heartbeat \cite{Pol-PhilMagJSci1928}.

The heartbeat instability was investigated in microgravity conditions on board the International Space Station \cite{Zhdanov-NewJPhys2010, Heidemann-AIPPhysPlas2011},
in experiments with grown nanoparticles \cite{Mikikian-AIPPhysPlas2004, Mikikian-NewJPhys2007, Mikikian-PhysRevLett2008, Mikikian-PhysRevLett2010}
and with the thermophoretic gravity compensation \cite{Pustylnik-AIPPhysPlas2012}.
These experiments showed that oscillations of the discharge electrical parameters and the plasma emission are inherent in the heartbeat instability.
Just before every void contraction, a short bright flash appears in the void.
In some experiments, small oscillations of the plasma emission and the electrical signals occurs between the void contractions 
\cite{Mikikian-NewJPhys2007, Mikikian-PhysRevLett2008}.
The heartbeat instability is sensitive to the amount and size of the microparticles, the gas pressure, and the discharge power
making it difficult to reproduce the instability quantitatively.
Such phenomena as ``trampoline effect'' \cite{Kretschmer-PhysRevE2005} and ``delivery instability'' \cite{Couedel-AIPPhysPlas2010}
seem to be qualitatively similar to the heartbeat instability.
In \cite{Pustylnik-AIPPhysPlas2012}, the heartbeat instability was resonantly stimulated by modulating the ionization rate with a tunable laser.
It remained, however, unclear, which oscillatory mode of the void was excited by the tunable laser.

The heartbeat instability represents a very complex process involving the microparticle-plasma interactions.
Despite several attempts to explain the heartbeat instability theoretically  \cite{Vladimirov-AIPPhysPlas2005, Zhdanov-NewJPhys2010, Pustylnik-AIPPhysPlas2012}, 
its physical mechanism is still unclear.
Understanding of this phenomenon is essential for the physics of microparticle-plasma interactions and the void formation.
It needs a comprehensive model of the microparticle dynamics coupled with plasma conditions.
For such a model, understanding of the void formation mechanism is required.

It is commonly supposed that the void in complex plasmas is formed due to a balance between the electrostatic and the ion drag forces 
with the ion flow from the void into the microparticle suspension \cite{Goree-PhysRevE1999, Vladimirov-AIPPhysPlas2005, Kretschmer-PhysRevE2005, Lipaev-PhysRevLett2007}.
However, this mechanism cannot explain the void formation at low discharge powers \cite{Pustylnik-PhysRevE2017, Pikalev-PlasmaSrcSciTech2021}.
The experiment and the model in \cite{Pikalev-PlasmaSrcSciTech2021} demonstrated 
that the void can exist in one of two regimes depending on the discharge power.
The so-called bright void regime corresponds to the conditions with higher RF discharge power 
and is characterized by the bright emission inside the void.
In this case, strong time-averaged electric field at the top and the bottom void boundaries heats up the electrons enhancing the ionization and the emission.
The mechanical stability of the boundary is in this case indeed determined by the balance of electrostatic and ion drag forces.
The dim void regime corresponds to relatively low discharge power conditions 
and is characterized by the diffuse emission profile inside and around the void.
The transition between the regimes has a discontinuous character.
After the void formation, its size is almost independent of the discharge power in the dim regime,
but the bright void grows rapidly with the discharge power increase after the transition.
A simplified time-averaged 1D fluid model in \cite{Pikalev-PlasmaSrcSciTech2021} could reproduce the dim void only when 
the radial ion losses were artificially introduced into the microparticle-free region of the plasma.
In this case, the force balance on the top and bottom void boundaries could be satisfied for low plasma density
and with much weaker time-averaged electric field.

Combining the comprehensive investigation of the optogalvanic control of the heartbeat instability with the concept of dim and bright void,
we decompose the entire heartbeat process into several stages, which can be investigated separately.

\section{EXPERIMENT}

\begin{figure*}[htb]
\centering
\includegraphics[width=1\textwidth]{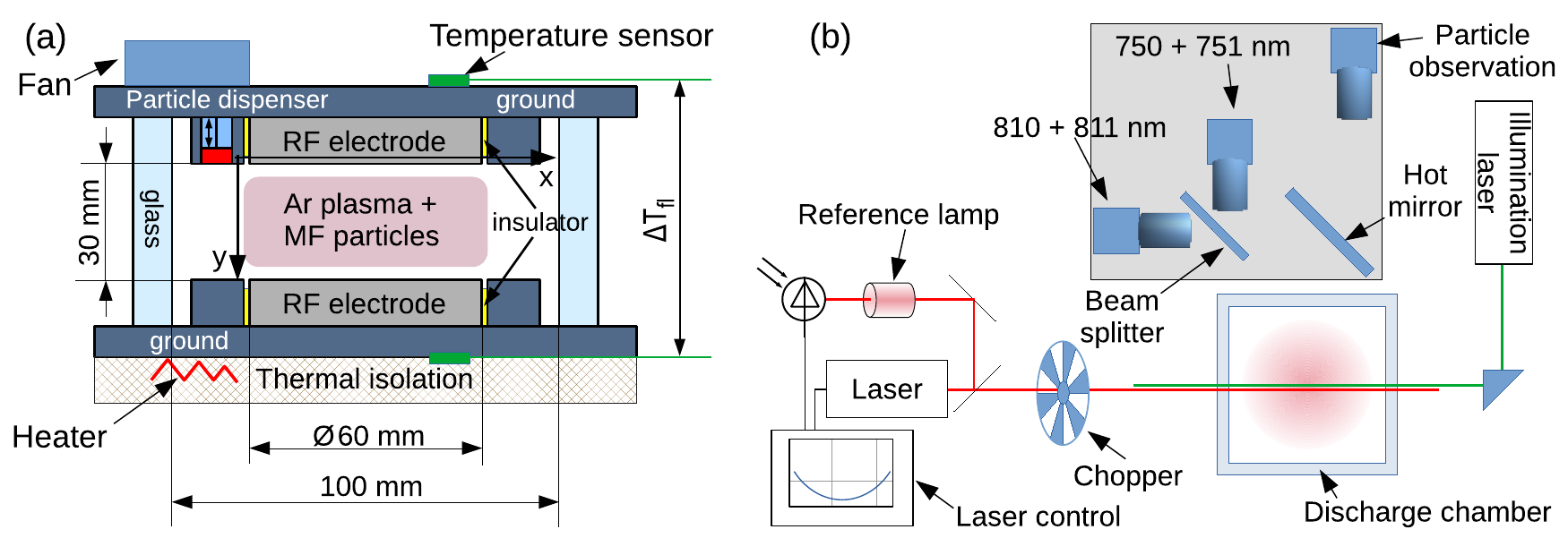}
\caption{Scheme of the experimental setup, (a) --- side view, (b) --- top view. 
The heart of the experiment is the PK-3 Plus chamber \cite{Thomas-NewJPhys2008}.
The bottom flange of the chamber can be heated to control the temperature gradient between the electrodes.
The cameras observe the microparticle motion and plasma emission.
The reference Ar lamp is used to control the wavelength of the tunable laser.
}
\label{scheme}
\end{figure*}

The experiments were conducted in the ground-based  PK-3 Plus chamber (see Fig. \ref{scheme}) \cite{Thomas-NewJPhys2008}. 
The plasma was produced by means of a capacitively coupled RF discharge. 
Two electrodes were driven in a push-pull mode by a sinusoidal signal with the frequency of 13.56~MHz.
We used two types of melamine formaldehyde spheres with the diameters of $1.95~\mu$m and $2.15~\mu$m as microparticles.
During a particular experiment, the microparticles of one of these types were injected into the discharge by an electromagnet-driven dispenser through a sieve.

Argon was fed into the chamber with 3~sccm gas flow during the entire experiment. 
We mainly investigated the heartbeat instability with the gas pressure of 36~Pa.
We also observed the heartbeat instability at the pressures of 20 and 55~Pa.
In the case of lower pressure, other instabilities \cite{Praburam-AIPPhysPlas1996, Samsonov-PhysRevE1999, Mikikian-IEEETPS2011}
made the discharge behavior complex and difficult to interpret.
In the case of higher pressure, gas friction made the heartbeat less pronounced.

\begin{figure}[htb]
\centering
\includegraphics[width=8.6cm]{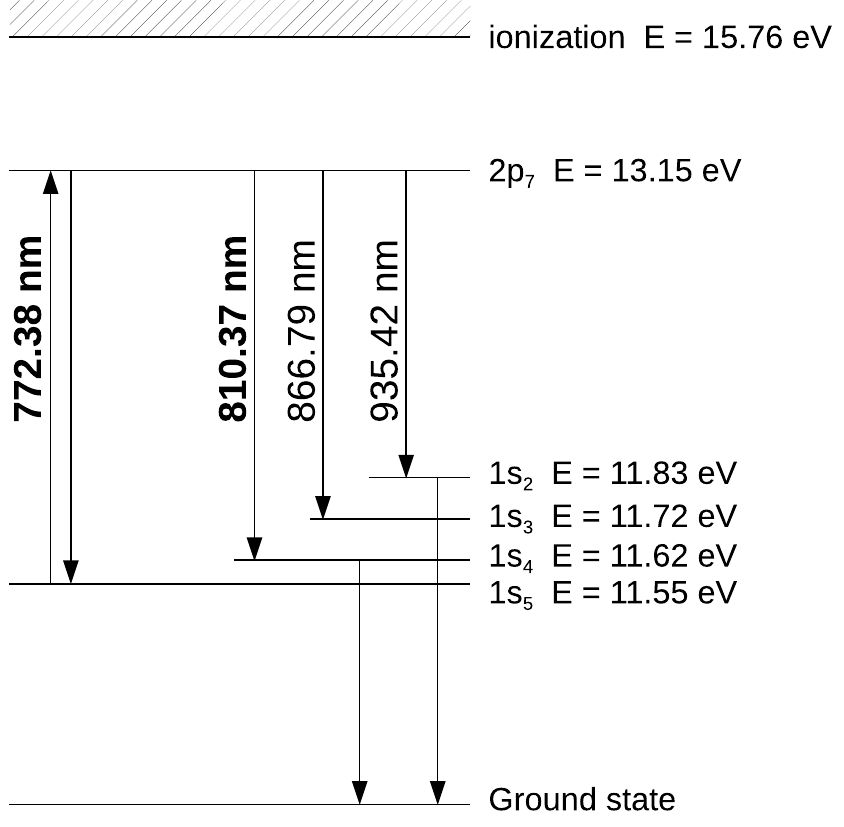}
\caption{Electronic levels of argon involved in the optogalvanic effect.
The laser pumps the transition at 772.38~nm and fluorescence in the 810.37~nm spectral line is observed.
}
\label{LIF-levels}
\end{figure}

Under ground laboratory conditions, the microparticles concentrate themselves in the vicinity of the lower electrode. 
To obtain large volumetric microparticle suspensions, we compensated the gravitational force by means of thermophoresis \cite{Rothermel-PhysRevLett2002}.
The temperature gradient between the electrodes was controlled by heating the bottom flange with an electric heater 
and cooling the top flange with fans.
A symmetrical void position was achieved with the temperature difference between the flanges of 14--15~K for the 1.95~$\mu$m diameter
microparticles and 15-17~K for the 2.15~$\mu$m diameter microparticles.

The microparticles were illuminated by a laser sheet with the wavelength of 532~nm.
Three Ximea MQ042RG-CM video cameras with interference bandpass filters 
captured the microparticle motion and the plasma emission in the same discharge area.
The filters had the central wavelengths of 532, 750, and 810~nm, respectively, and the width of transmission band of the filters was about 10~nm.
Hence, one plasma glow observation camera captured the plasma radiation in 750.4 and 751.5~nm spectral lines,
these lines correspond to the transitions 2p$_{1}$\,$\rightarrow$\,1s$_{2}$ and 2p$_{5}$\,$\rightarrow$\,1s$_{4}$ in the Paschen notation, respectively.
Another camera captured the light in 810.4 and 811.5~nm spectral lines,
which correspond to the transitions 2p$_{7}$\,$\rightarrow$\,1s$_{4}$ and 2p$_{9}$\,$\rightarrow$\,1s$_{5}$, respectively.
The spatial resolution of the cameras was about 37~$\mu$m/pixel.
The framerate of the cameras was 200~fps.
The video capture of all three cameras was started by the software approximately at the same time.
The difference between the timestamps of the frames with the same number was less than the duration of a single frame.
All three cameras were focused on the central cross-section of the discharge chamber.

We used the optogalvanic effect \cite{Goldsmith-ContempPhys1981,Barbieri-RevModPhys1990}
induced by the Toptica DL Pro laser for the investigation of the void stability \cite{Pustylnik-AIPPhysPlas2012}. 
The width of the laser spectral line was less than 1~MHz.
The power of the laser beam entering the plasma chamber was about 50~mW, and its diameter was 2~mm.
The laser was tuned to the 772.38 nm transition of argon and scanned the spectral range of about 220 MHz in the vicinity of its center with the repetition frequency of 100~Hz.
A small fraction of the laser light passed though a reference argon lamp to control the scanning range. 
A mechanical chopper was used to modulate the laser beam.
The laser light induced the fluorescence in the 810.4~nm spectral line (see Fig. \ref{LIF-levels}),
which was used to match the laser modulation with the video frames.

At the beginning of every experiment, after stabilization of the temperature gradient, we set the discharge power to 500~mW 
and injected the microparticles till the onset of the heartbeat instability.
The parameters of the instability, such as its frequency, the threshold discharge power of the instability onset, and
necessary laser parameters for the optogalvanic control,  
drifted on the timescales of minutes making the measured quantities difficult to reproduce. 
Such evolution of the conditions is typical for complex plasma experiments \cite{Killer-PlasmaSrcSciTech2016}.
However, the qualitative effects reported in this paper are well reproducible.

\section{RESULTS}

\subsection{Self-excited instability}

The self-excited heartbeat instability existed within a certain range of the discharge power and the gas pressure 
and required relatively high density of the microparticle suspension.
An example of the heartbeat self-excitation region in pressure - discharge power space was published in \cite{Pustylnik-AIPPhysPlas2012}.
This region was expanding with the increase of the amount of microparticles.

At the lower discharge power boundary of the instability range, the void underwent sporadic contractions with strong variation of the void size,
while near the higher power boundary, the void size variation was decreasing with the increase of the discharge power, 
until the oscillations became undetectable.
For example, in one experiment with the microparticle diameter of 2.15~$\mu$m, the heartbeat appeared as the sporadic contractions at the discharge power of 500~mW
with the variation of the void cross-section area between 31 and 7~mm$^2$.
After the discharge power was increased to 750~mW,
the void cross-section area oscillated between 30 and 27~mm$^2$ with the frequency of 26~Hz.
In other our experiments, the lower discharge power of the heartbeat self-excitation was 300--600~mW, 
and the oscillations became invisible for the discharge powers higher than 800~mW.

The reported experiments were conducted near the lower power boundary, where the instability onset had the threshold character and 
the void oscillations were well pronounced.
In the experiments with the self-exited heartbeat instability, 
the amount of microparticles only slightly exceeded the instability onset threshold.

\begin{figure*}[htb]
\centering
\subfloat{
\includegraphics[width=0.4\textwidth]{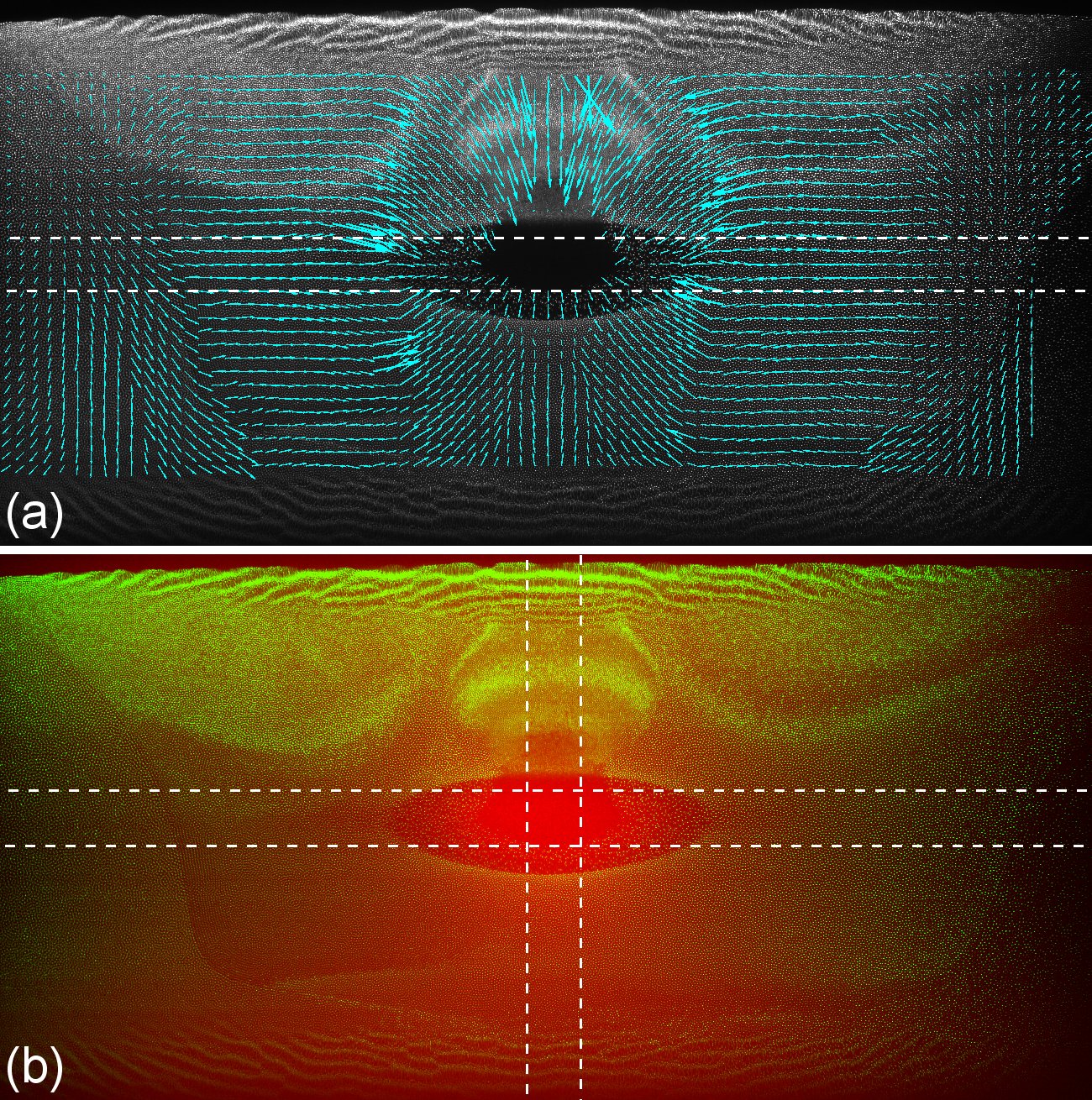}
}
\hfill
\subfloat{
\includegraphics[width=0.56\textwidth]{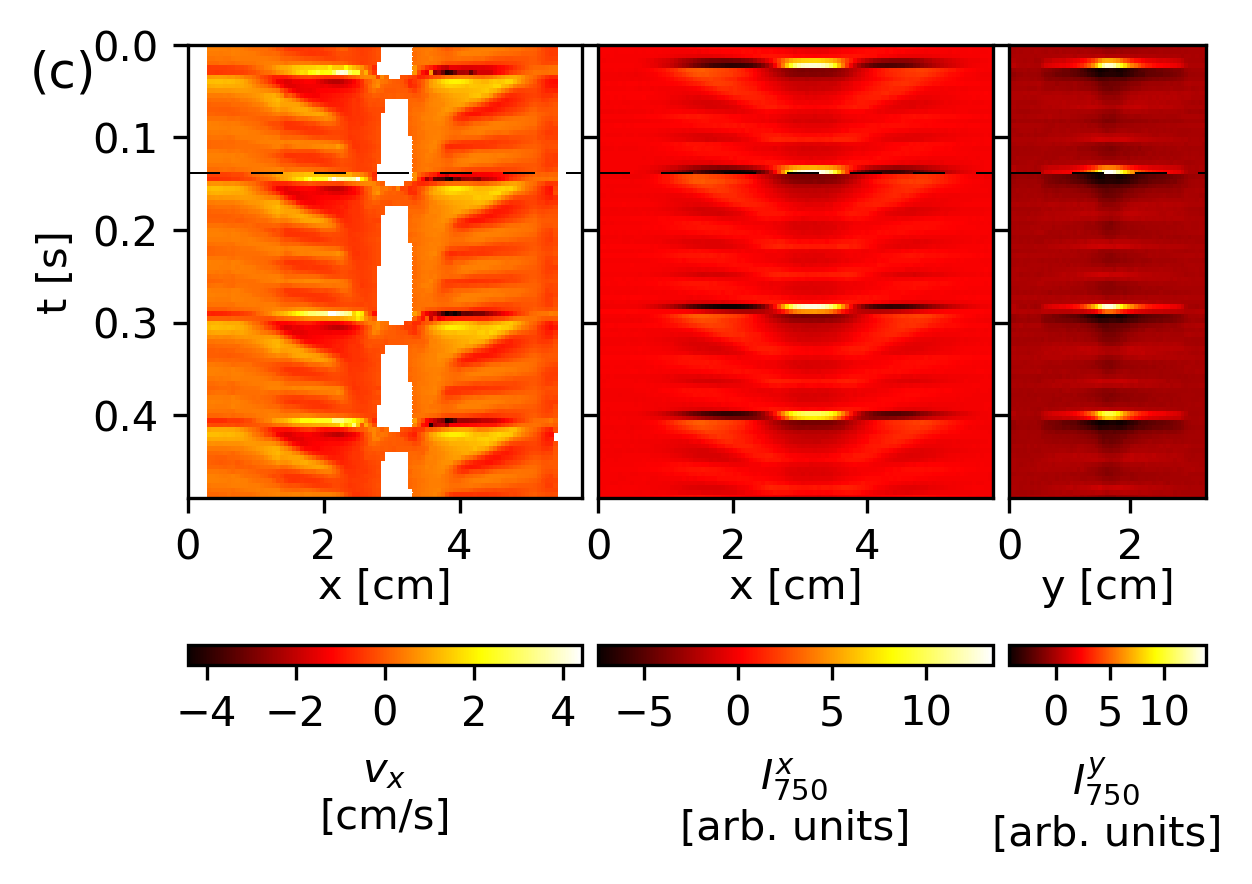}
}
\caption{
Self-excited heartbeat instability.
(a) One of the frames with the velocity field reconstructed using the OpenPIV software \cite{Taylor-IEEETransInstrumMeasur2010, *openPIV-qt-github}.
(b) The same frame with superimposed images of (green) the microparticle suspension
and (red) the plasma emission captured with the filter with the central wavelength of 750~nm.
The dashed lines in plates (a) and (b) depict the areas used for the calculation of the spatiotemporal distributions.
(c) Spatiotemporal distributions of the velocity $v_x$ and 
the plasma emission variations $I_{750}^x$ and $I_{750}^y$ during the self-excited heartbeat instability. 
The white color in the $v_x$ plot depicts the areas in which no microparticles are present. 
The black dashed lines depict the temporal position of the frames shown in plates (a) and (b).
The microparticle diameter was 2.15~$\mu$m, the discharge power was 400~mW.
See also supplemental materials \cite{supplemental}.
}
\label{self_excited}
\end{figure*}

Similarly to \cite{Zhdanov-NewJPhys2010}, we obtained the spatiotemporal distributions (periodograms) of the microparticle velocity.
First, we calculated the velocities $v_x$ using OpenPIV code \cite{Taylor-IEEETransInstrumMeasur2010, *openPIV-qt-github}
in cells 32x32 pixels with 50\% overlap.
Then, for every frame, we averaged the velocities along the vertical direction in a horizontal stripe with the width of four cells.
The stripe was located at the height of the void [see Fig. \ref{self_excited}(a)]. 
The obtained horizontal velocity profiles were stacked into a spatiotemporal distribution [see an example in Fig. \ref{self_excited}(c)].

In a similar way, $x$ and $y$ distributions of emission intensity  were obtained 
by stacking the profiles of the averaged emission intensities across the respective narrow stripes.
The directions of $x$ and $y$ axes are shown in Fig. \ref{scheme}(a).
In order to obtain the emission variation, the time-averaged intensity was subtracted.
The emission variations for the filter with the central wavelength of 750 nm in horizontal and vertical directions
($I_{750}^x$ and $I_{750}^y$, respectively) are presented in Fig. \ref{self_excited}(c).
Just before the void contraction, the plasma glow shortly flashed in the void 
[see Fig. \ref{self_excited}(b, c) and movie\_fig3.avi in supplemental materials \cite{supplemental}]. 
The intensity at the flash maximum was 30--40\% higher than the time-averaged value.
At the same time, the discharge edge regions became darker.
After that, the emission decreased in the contracted void, 
whereas the emission from the edge regions increased above the level before the flash.
This effect was already reported in several works \cite{Mikikian-AIPPhysPlas2004, Mikikian-NewJPhys2007, Mikikian-PhysRevLett2010, Pustylnik-AIPPhysPlas2012}.
The plasma emission measured with 750 and 810~nm filters behaved similarly.
If the amount of microparticles in the suspension was slightly below the instability onset threshold, 
the spatial distribution of plasma emission was diffuse. 
Therefore, in this case, the void was in the dim regime (in the terminology of \cite{Pikalev-PlasmaSrcSciTech2021}).

Small oscillations of the microparticle motion and the plasma emission 
are visible in the spatiotemporal distributions between the large void collapses.
They were termed ``failed contractions'' in \cite{Mikikian-NewJPhys2007, Mikikian-PhysRevLett2010}.
In the velocity spatiotemporal distributions, a small slope of the oscillations can be seen,
which suggests that the oscillation propagated from the discharge edges to the center.
In our paper, we would therefore refer to them as to ``breathing oscillations''.

\subsection{Optogalvanic control}

We will describe the effects of the optogalvanic control of the heartbeat instability using the three spatiotemporal distributions: 
the distributions of $v_x$ and $I_{750}^x$ as in Fig. \ref{self_excited}(c) 
and additionally $x$-distribution of the emission intensity variation for the filter with the 810 nm central wavelength $I_{810}^x$. 
The $I_{810}^x$ distribution shows the periods when the laser was open or closed in the experiments with the modulated laser 
due to the laser-induced fluorescence (see Fig.~\ref{LIF-levels}).
In all cases, these distributions were obtained in the same way using the technique described in Sec 3.1.

\subsubsection{Continuous laser}

\begin{figure}[htb]
\centering
\includegraphics[width=8.6cm]{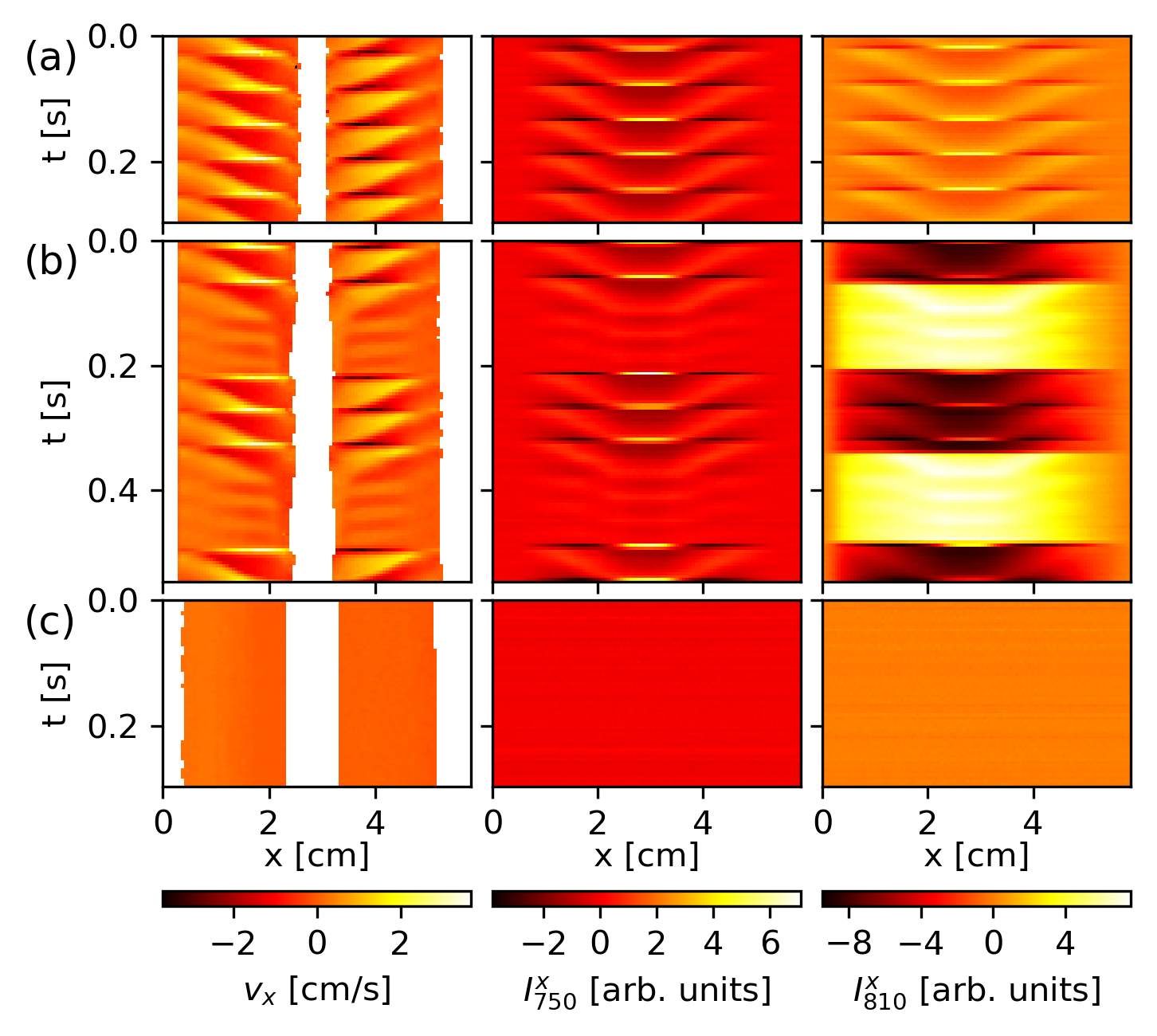}
\caption{
The horizontal spatiotemporal distributions of the velocities $v_x$ and
the plasma emission variations $I_{750}^x$ and $I_{810}^x$ (a) without the laser, (b) with the modulated laser, and (c) with continuous laser.
The diameter of the microparticles was 1.95~$\mu$m, the discharge power was 500~mW.
See also supplemental materials \cite{supplemental}.
}
\label{stabilization}
\end{figure}

The continuous laser beam passing through the void stabilized the microparticle suspension.
Without the laser, the heartbeat instability was present in the suspension.
The frequency of the heartbeat instability in that case was 17.9 Hz [see Fig.~\ref{stabilization}(a)].
We modulated the laser with the frequency of 3.6 Hz. 
While the laser was closed, the heartbeat instability was observed in the suspension. 
However, once the laser was opened, 
the instability disappeared and only breathing oscillations being slowly damped remained visible.
This effect is demonstrated in Fig. \ref{stabilization}(b) and movie\_4b.avi in supplemental materials \cite{supplemental}.
The continuous laser beam stabilized the void completely, and no oscillations were visible [see Fig.~\ref{stabilization}(c)].

\subsubsection{Transient stimulation}
\begin{figure}[htb]
\centering
\includegraphics[width=8.6cm]{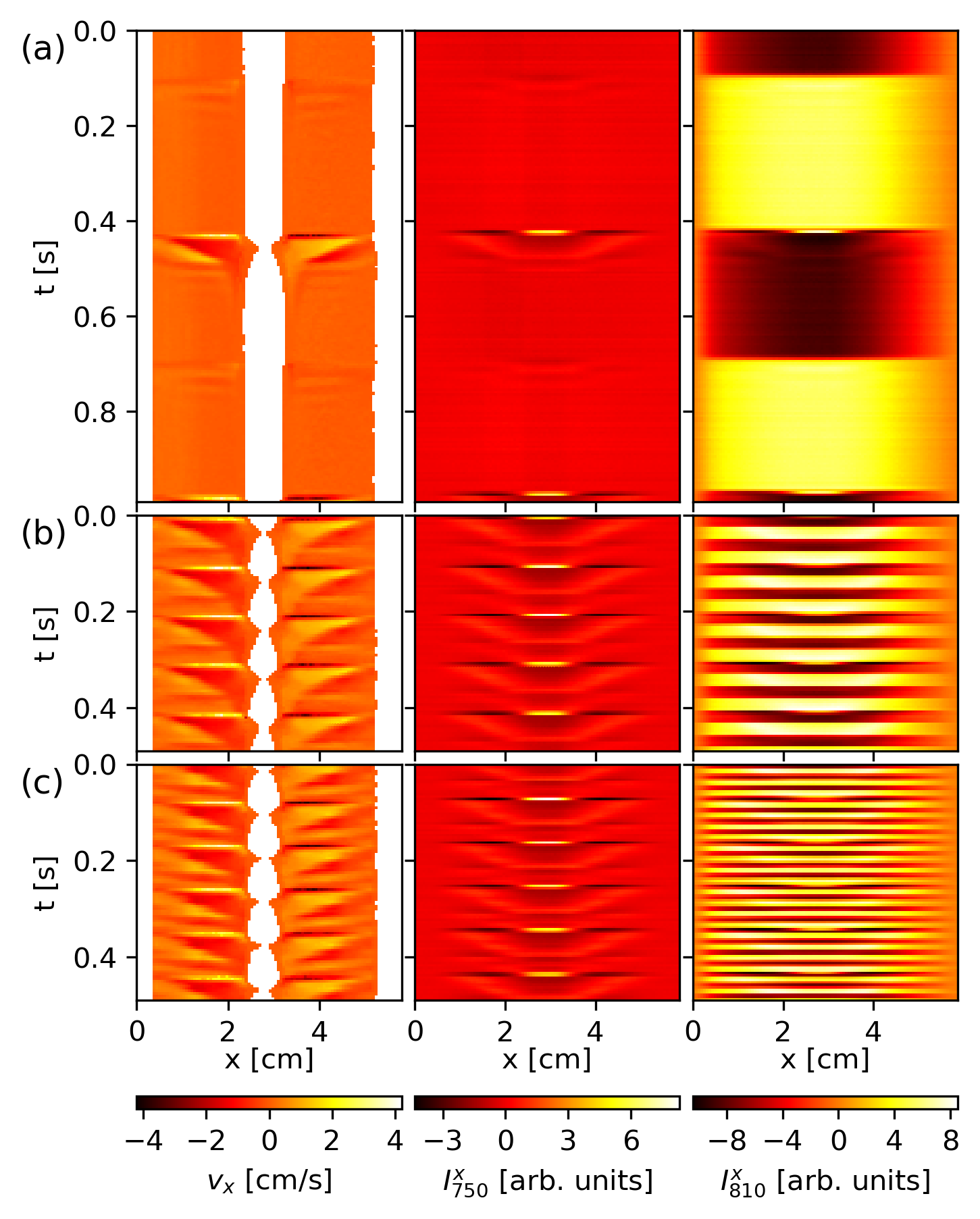}
\caption{Transient optogalvanic heartbeat excitation with the chopper frequency of (a) 1.7~Hz, (b) 19.6~Hz, (c) 44.0~Hz.
The microparticle diameter was 1.95~$\mu$m, the discharge power was 500~mW.
See also supplemental materials \cite{supplemental}.
}
\label{non-resonant}
\end{figure}

The laser beam could transiently, either on its opening or closing, excite the heartbeat instability
in a stable microparticle suspension.
If the laser beam passed through the void, 
the void collapse occurred just after the beam closing.
The spatiotemporal distributions of the transient heartbeat stimulation are presented in Fig. \ref{non-resonant} 
and corresponding movies are available in supplemental materials \cite{supplemental}.
In the case of the chopper frequency of 1.7~Hz [Fig. \ref{non-resonant}(a)], every closing of the beam caused the void collapse. 
Also, small breathing oscillations quickly damped after opening the beam are visible.
For the chopper frequencies of 19.6 Hz [Fig. \ref{non-resonant}(b)] and 44.0 Hz [Fig. \ref{non-resonant}(c)], 
the void collapses occurred every two and every four laser pulses, respectively.
It seems, the void needs to reach a certain phase of expansion after the collapse to be ready for a new stimulated contraction.

At the same time, the chopper frequency must not be too low.
In one experiment, the transiently induced contractions occurred with the chopper frequency of 4.6~Hz,
but did not appear with the chopper frequency of 3~Hz.

\begin{figure*}[htb]
\centering
\includegraphics[width=1\textwidth]{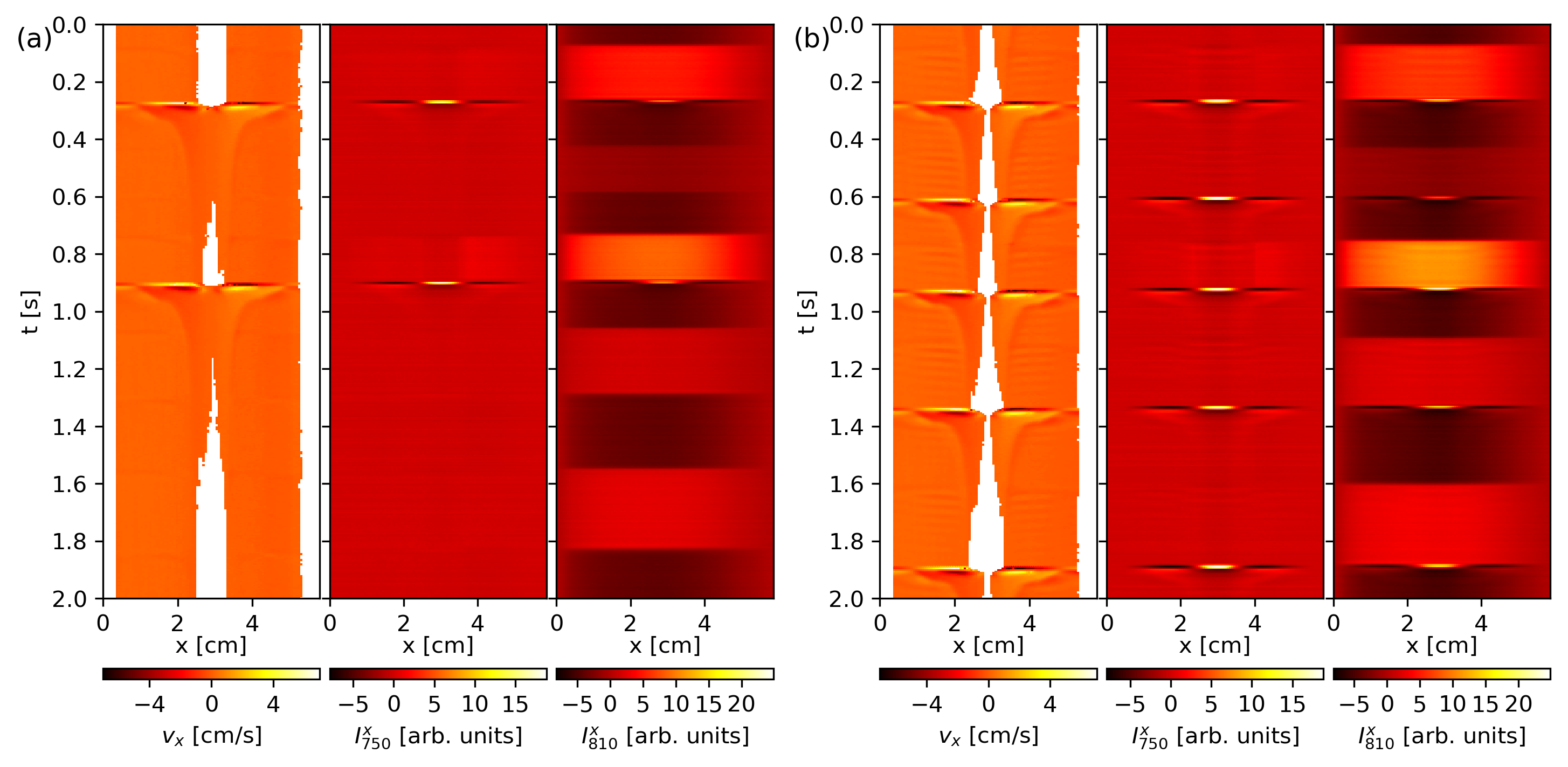}
\caption{Transient optogalvanic heartbeat excitation with the attenuated laser beam.
The microparticle diameter was 1.95~$\mu$m, the discharge power was (a) 250~mW, (b) 350 mW.
The bright stripes in the $I^x_{810}$ distribution correspond to the time intervals 
when the laser was crossing the void through the filters with the following values of the transmission (from the top to the bottom): 0.45, 0.014, 1, 0.12, and 0.19.
}
\label{filters}
\end{figure*}

To evaluate how the effect depends on the laser intensity, we equipped the chopper with neutral density filters of different transmission. 
The experiments showed that the further the suspension is from the self-excitation threshold, 
the higher laser power is necessary to stimulate the contraction.
An example of the results is presented in Fig. \ref{filters}.
In this experiment, the measurements with the discharge power of 250~mW were performed 6 minutes after 
the measurements with the discharge power of 350~mW without injection of the microparticles in between.

\begin{figure}[htb]
\centering
\includegraphics[width=8.6cm]{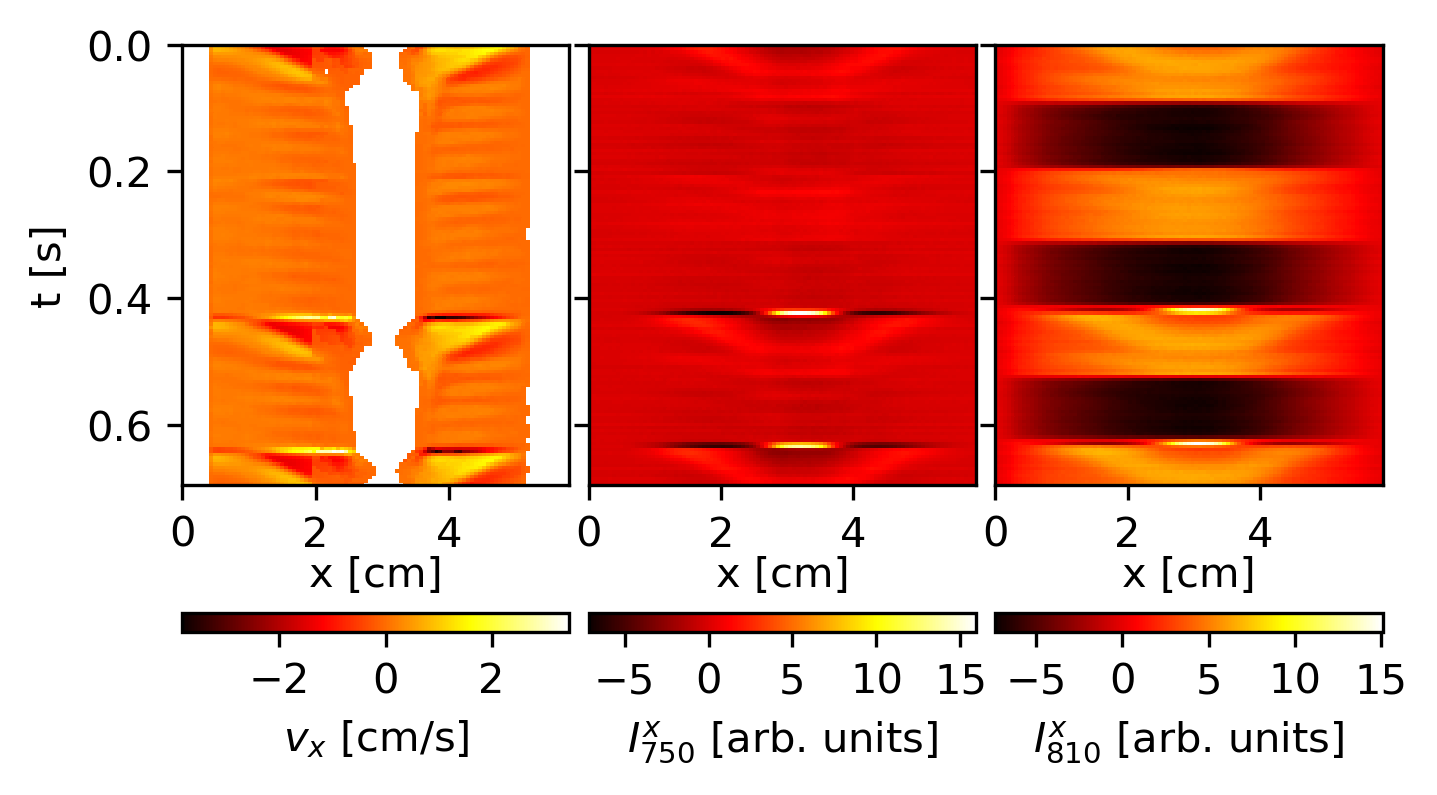}
\caption{Effect of the beam horizontal shift by 10~mm.
The microparticle diameter was 2.15~$\mu$m, the discharge power was 500~mW.
}
\label{shift}
\end{figure}

If the laser beam was shifted horizontally from the void center to the distance of 1~cm, 
the void collapsed after the laser opening (see Fig. \ref{shift}) in contrast with the experiments 
where the unshifted laser beam caused the void contraction after closing.
We could not obtain this change of the void collapse phase with a vertical shift of the laser beam:
the effect of the laser beam under or above the void became weaker, as it was reported in \cite{Pustylnik-AIPPhysPlas2012},
and only rare void contractions occurred, still after the laser closing.
This difference between the horizontal and vertical shifts of the beam corresponds to the difference 
between horizontal and vertical spatiotemporal emission distributions during the flash [see Fig~\ref{self_excited}(c)].

\subsubsection{Resonant stimulation}

\begin{figure}[htb]
\centering
\includegraphics[width=8.6cm]{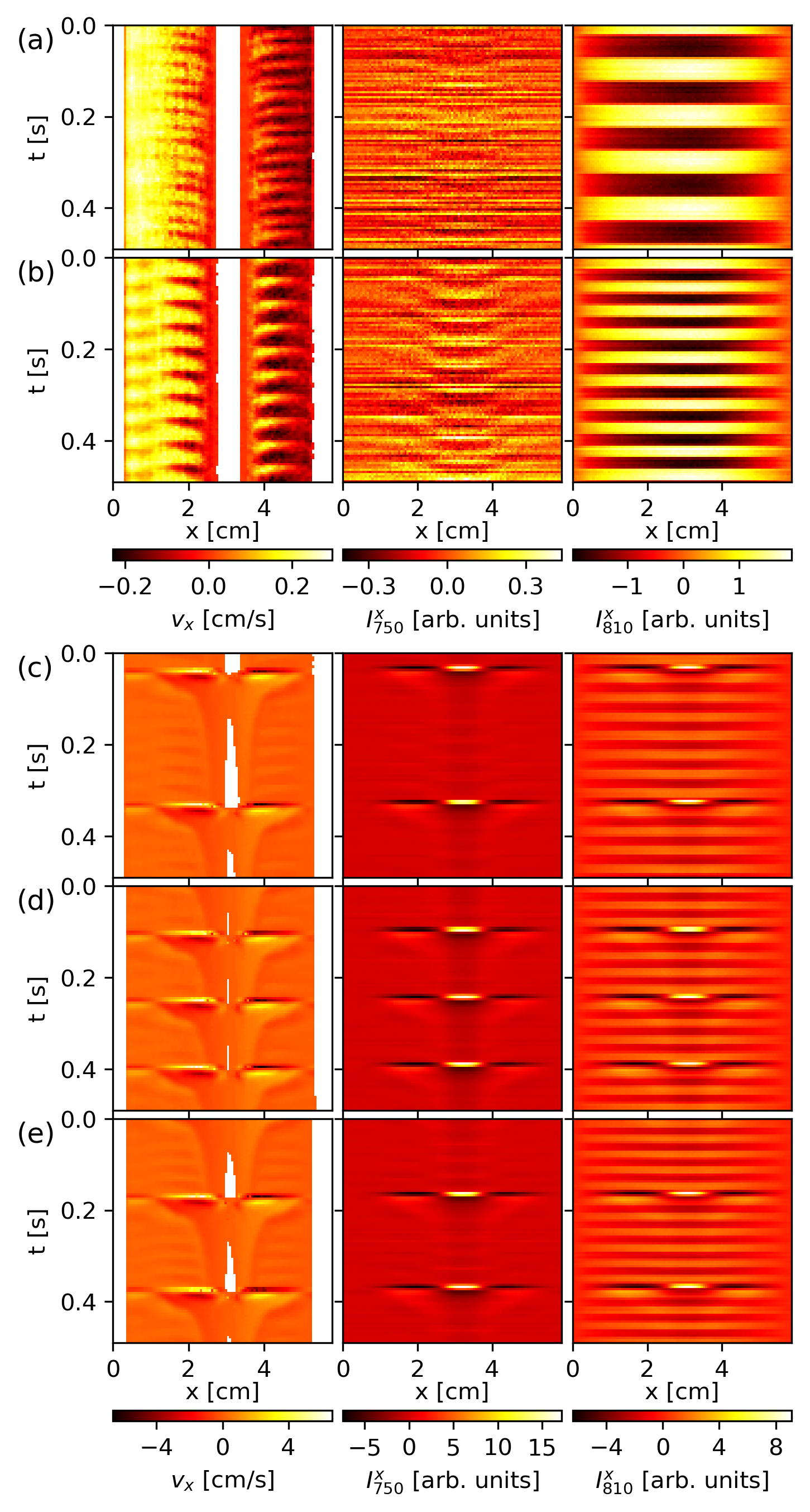}
\caption{Stimulation of the heartbeat instability with low laser power.
The laser beam was attenuated down to 0.7~mW.
The microparticle diameter was 2.15~$\mu$m, the discharge power was 350~mW.
The laser modulation frequency was (a) 9.8~Hz, (b) 19.5~Hz, (c) 23.9~Hz, (d) 27.3~Hz, and (e) 29.5~Hz.
See also supplemental materials \cite{supplemental}.
}
\label{resonant}
\end{figure}

\begin{figure}[htb]
\centering
\includegraphics[width=8.6cm]{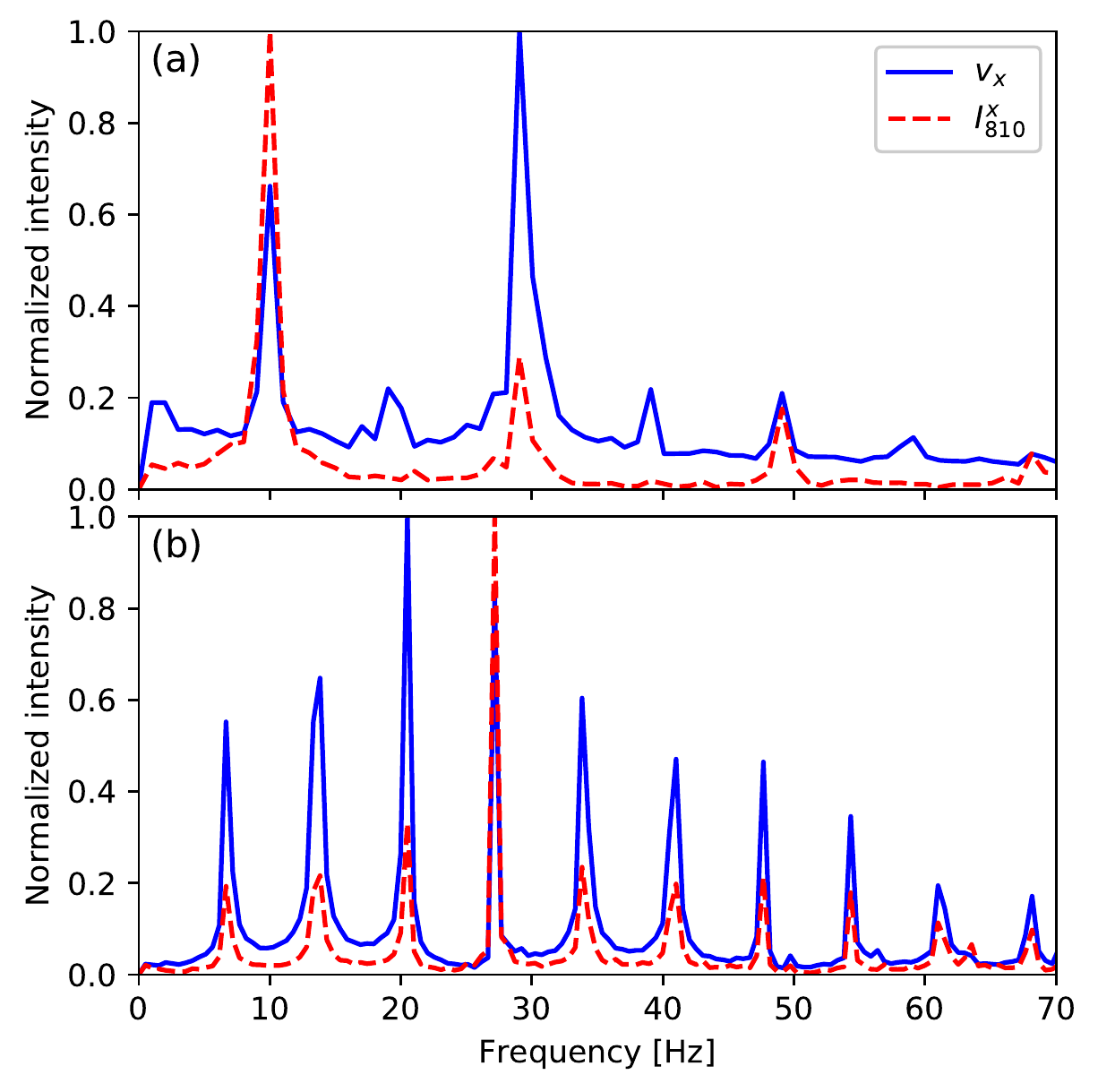}
\caption{ Horizontally averaged Fourier spectra of $v_x$ and $I^x_{810}$ spatiotemporal distributions 
shown in (a) Fig. \ref{resonant}(a); (b) Fig. \ref{resonant}(d).
The laser modulation frequency was (a) 9.8~Hz, (b) 27.3~Hz.
Every spectrum is normalized to its maximal value.
}
\label{spectrum}
\end{figure}

If the void contractions could not be induced transiently,
the heartbeat instability could be stimulated resonantly.
This effect was reported in \cite{Pustylnik-AIPPhysPlas2012} for the first time.
It was discovered using the same laser and discharge chamber as in this work.
An example of this effect is shown in Fig. \ref{resonant}.
In the case of the lowest chopper frequency [Fig. \ref{resonant}(a)], small breathing oscillations,
whose phase is disturbed by the modulated laser beam, are visible.
Horizontally averaged Fourier spectra of the spatiotemporal distributions for this case are shown in Fig. \ref{spectrum} (a).
In this spectrum, mainly two modes are visible: 
The first one with the frequency of about 10~Hz corresponds to the laser modulation, 
whereas the second one with the frequency of about 29~Hz corresponds to the breathing oscillations.
At higher chopper frequency, the microparticle suspension oscillated 
with the frequency of the laser modulation and low amplitude [Fig. \ref{resonant}(b)].
At the chopper frequencies close to the frequency of the breathing oscillations,
the heartbeat instability occurred [see Fig. \ref{resonant} (c--e) and movie\_fig8d.avi in supplemental materials \cite{supplemental}].
The width of this resonance was several Hz.

The Fourier spectra for the spatiotemporal distributions in Fig.~\ref{resonant}(d) are shown in Fig.~\ref{spectrum}(b).
Mixing of the modulation frequency of 27.3~Hz with four times lower heartbeat frequency of 6.8~Hz leads to the frequency comb spectrum \cite{Holzwarth-PhysRevLett2000}.
We note that the spectrum of the heartbeat instability observed in \cite{Mikikian-AIPPhysPlas2004, Mikikian-PhysRevLett2008} also has a frequency comb character.

\section{DISCUSSION}

\subsection{Dim-to-bright void transition and void collapse}

In \cite{Pustylnik-AIPPhysPlas2012}, the heartbeat instability was supposed to occur due to a critical transformation on the void boundary
from a smooth potential profile to the sheath.
The sheath was suggested to appear when the microparticle number density reaches a certain value:
When the amount of microparticles is small, they do not significantly disturb the plasma, and the potential profile across the void boundary should be smooth.
If the suspension is however dense, the void boundary should act similarly to a solid surface, in front of which the sheath potential drop should occur.
However, simulation with a 1D hybrid model and fixed microparticle distribution showed 
a linear growth of the electric field with increasing the microparticle number density without any critical phenomena \cite{Pustylnik-APCPTS2018}.
According to recent findings \cite{Pikalev-PlasmaSrcSciTech2021}, a steep growth of the electric field accompanies 
the transition between dim and bright void regimes.
Since the heartbeat instability is accompanied by the flash inside the void, 
it is reasonable to suppose that the void contractions are caused by an abrupt transition between dim and bright void regimes.

This process is fast in comparison with the microparticle motion:
In \cite{Pustylnik-AIPPhysPlas2012} the flash was measured with a photomultiplier 
and looked as a pulse with the duration of 5~ms and steep edges.
In the experiments with growing nanoparticles, the flash duration of 2~ms was reported \cite{Mikikian-NewJPhys2007}.
It leads therefore to the transient formation of a bright-void-like emission (and, consequently, ionization) profile under the plasma conditions of a dim void.   
Similar situation was considered in \cite{Pustylnik-PhysRevE2017} using a 1D hybrid model:
simulations with a fixed microparticle distribution with a void resulted in a strong emission inside the void 
and uncompensated electrostatic forces pushing the microparticles to the void center.

\subsection{Optogalvanic effect}

In \cite{Pikalev-PlasmaSrcSciTech2021}, it was pointed out that an increase of the ionization rate may switch the void from the dim to bright regime.
The laser whose wavelength is tuned to the electron transition of the plasma forming gas atoms is capable of modifying the ionization rate in the plasma.
This is the underlying mechanism of the optogalvanic effect \cite{Nestor-ApplOpt1982, Murnick-ApplPhysLett1989}.
The ionization energy of argon from the ground state is 15.76~eV \cite{Minnhagen-JOptSocAm1973},
which is much higher than the electron temperature in our plasma ($T_e \sim 4\,$eV, $n_e \sim 10^{15}\,$m$^{-3}$ \cite{Pustylnik-PhysRevE2017, Pikalev-PlasmaSrcSciTech2021}).
The bottom state of the transition pumped by the laser is the lowest metastable level,
which lies only 4.21~eV below the ionization energy.
Although, after additional excitation by the laser, the atom can be ionized even easier,
it can spontaneously relax to the ground state (see Fig. \ref{LIF-levels}).
The optogalvanic effect is therefore determined by the interplay between the growth of the ionization rate 
due to the increase in the population of the upper level of the laser-pumped transition 
and the decrease of the ionization due to the decrease in the population of the lower level of the laser-pumped transition.
The sign of the effect depends on the experimental conditions. 
For example, nanosecond laser pulses in the 696.5~nm argon spectral line caused a decrease of ionization in a glow discharge \cite{Nestor-ApplOpt1982},
whereas much longer pulses (0.4~ms) in the same spectral line increased the ionization in a RF discharge \cite{Murnick-ApplPhysLett1989}. 

In our case, the probability of the relaxation to $1s_2$ and $1s_4$ resonant states is one order of magnitude higher
than the probability to relax to the $1s_3$ metastable state \cite{Wiese-PhysRevA1989}. 
According to \cite{Nestor-ApplOpt1982}, a laser pulse in 772.38~nm argon spectral line
caused a positive pulse of DC discharge voltage (at constant discharge current),
which implies the decrease of the ionization rate.
The laser-induced decrease of the ionization rate is consistent 
with the stabilization of the heartbeat instability by the continuous laser since the dim void condition corresponds to the lower ionization rate \cite{Pikalev-PlasmaSrcSciTech2021}.

\subsection{Laser control of the heartbeat instability}

In the case of self-excited heartbeat instability, the transition from the dim to bright void regime is evidently mediated by the breathing oscillations: 
They are always seen as precursors of the heartbeat instability as pointed out also in \cite{Mikikian-NewJPhys2007, Mikikian-PhysRevLett2008}.
The dim void formation requires significant radial ion losses \cite{Pikalev-PlasmaSrcSciTech2021}, which could be modulated by the breathing oscillations. 
In the case of resonant optogalvanic stimulation of the heartbeat instability, the laser slightly modulates the ionization rate (as pointed out in \cite{Pustylnik-AIPPhysPlas2012}). 
The heartbeat instability is excited when the modulation frequency is close to the frequency of the breathing oscillations. 
Also, the breathing oscillations are damped, once the heartbeat instability is switched off by the continuous laser beam [see Fig.~\ref{stabilization}(c)].

During the transient heartbeat stimulation, the transition from dim to bright void occurs immediately after opening or closing the laser beam
independent of the existence or phase of the breathing oscillations.
We suppose that in this case, the ionization rate inside the void increases due to the transient processes in the plasma.
If the laser beam passes though the void, its closing causes this increase directly.
If the beam is shifted, a decrease of the ionization near the beam requires an ionization increase in other parts of the discharge.

\section{CONCLUSION}

\begin{figure}[htb]
\centering
\includegraphics[width=8.6cm]{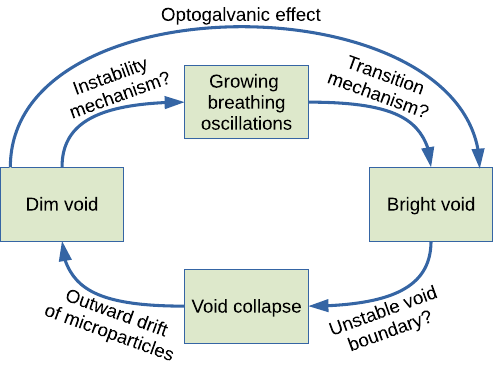}
\caption{\label{cycle} Schematic representation of the heartbeat instability cycle.
Initially dim void transits to the bright regime. 
This transition is mediated either by the breathing-oscillations-related mechanism or by the transient optogalvanic effect.
The boundary of the bright void becomes unstable and the void collapses due to the electrostatic force prevailing over the ion drag and thermophoretic forces.
The dim void recovers due to the outward drift of microparticles.}
\end{figure}

The presented experiments reproduced earlier reported behavior of the plasma emission during the heartbeat instability,
small breathing oscillations that appear in the suspension between the void contractions,
and resonant optogalvanic stimulation of the instability.
It was demonstrated that the resonance is observed when the laser modulation frequency coincides with the frequency of these breathing oscillations.
The experiments with high laser power revealed that individual laser switching can cause the void contraction transiently.
The void contracts after closing of the laser beam if it passes through the void center,
and after opening of the laser beam if  it was shifted horizontally to the void periphery.
However, continuous laser beam stabilizes the microparticle suspension.

The experimental results allow us to describe the cycle of the heartbeat instability in the following way (Fig. \ref{cycle}): 
It starts from a microparticle suspension with a dim void. 
The void transits abruptly to the bright regime due to the breathing oscillations, whose amplitude grows beyond a critical value \cite{Mikikian-NewJPhys2007}.
The transition can also be induced by a steep change of ionization rate by the laser or the laser working in resonance with the breathing oscillations.
In the bright regime, the void boundary becomes mechanically unstable and the void collapses due to large electrostatic forces on its boundaries.
After the collapse, the microparticles absorb the surplus of electrons and ions and move back restoring the dim-void configuration.

We were, therefore, able to identify a sequence of physical conditions the plasma acquires during the cycle of the heartbeat instability (green boxes in Fig. \ref{cycle}). 
However, most of the transitions between them (arrows in Fig. \ref{cycle}) are still not clear.
What is the self-excitation mechanism of the breathing oscillations in the dim void regime? 
How does their growth lead to the abrupt transition to the bright void regime? 
Why does the void boundary in the bright regime get mechanically unstable? 
What makes the microparticles move backward and form the dim void after the bright void collapse?
In the discussion we suggested hypothetical mechanisms which could explain some of these transitions.
They need to be proved by further investigations, theoretical as well as experimental.

As we already mentioned, the heartbeat instability is a multi-time-scale process. 
Its direct simulation is, therefore, computationally very expensive.
Decomposing the heartbeat into stages allows to model each of them individually.
The optogalvanic control allows to experimentally separate the void collapse from the breathing oscillations.
This approach could be applied to other instabilities in complex plasmas.

\begin{acknowledgments}
We would like to thank Dr. V. Nosenko for careful reading of our manuscript.
The PK-3 Plus chamber was funded by the space agency of DLR 
with funds from the federal ministry for economy and technology according to a resolution of the Deutscher Bundestag under grants No. 50WP0203, 50WM1203.
A.~Pikalev acknowledges the financial support of Deutscher Akademischer Austauschdienst (DAAD) with funds from 
Deutsches Zentrum f\"{u}r Luft- und Raumfahrt e.V. (DLR). 
\end{acknowledgments}

\bibliography{pikalev_heartbeat}

\newpage

\onecolumngrid

\section*{SUPPLEMENTAL MATERIALS}
The supplemental materials contain videos with the data that were used to obtain some of the results presented in the paper.
The video frames are constructed by superimposing colour-coded frames of the microparticle suspension (green) 
and the plasma emission captured through the filters with 750 nm (red) and 810 nm (blue) central spectral line.
The framerate is decreased to 10 fps making the videos 20 times slower than the real processes.

\begin{description}
\item[movie\_fig3.avi] Self-exited heartbeat instability. See Fig. 3 for details.

\item[movie\_fig4b.avi] Stabilization of the microparticle suspension during the periods with open laser. See Fig. 4(b) for details.

\item[movie\_fig5a.avi, movie\_fig5b.avi, movie\_fig5c.avi] Transient optogalvanic heartbeat excitation with the chopper frequency of 
1.7 Hz, 19.6 Hz, and 44.0 Hz, respectively. See Fig. 5 for details.

\item[movie\_fig8d.avi] Resonant heartbeat stimulation. See Fig. 8(d) for details.
\end{description}

Some very small particles, which are visible near the upper void boundary in the videos, seem to be a contamination of the injected microparticles.
They appear right after an injection.
These smaller particles have much faster reaction.
However, in the observations presented in the manuscript, the heartbeat frequency was so low
that both the normal microparticles and the smaller particles behave qualitatively similar.

\end{document}